\documentclass{article}

\usepackage{iclr2027_conference,times}

\usepackage{graphicx}
\usepackage{my/myconfig}
\usepackage{hyperref}
\usepackage{url}

\definecolor{cellcolor}{RGB}{0,0,0}
\definecolor{antiquewhite}{rgb}{0.98, 0.92, 0.84}
\definecolor{anti-flashwhite}{rgb}{0.95, 0.95, 0.96}
\definecolor{aliceblue}{rgb}{0.94, 0.97, 1.0}
\definecolor{almond}{rgb}{0.94, 0.87, 0.8}
\definecolor{cosmiclatte}{rgb}{1.0, 0.97, 0.91}
\definecolor{darkbyzantium}{rgb}{0.36, 0.22, 0.33}
\definecolor{darkseagreen}{rgb}{0.56, 0.74, 0.56}
\definecolor{darkspringgreen}{rgb}{0.09, 0.45, 0.27}
\definecolor{asparagus}{rgb}{0.53, 0.66, 0.42}
\definecolor{antiquefuchsia}{rgb}{0.57, 0.36, 0.51}
\definecolor{ao(english)}{rgb}{0.0, 0.5, 0.0}
\definecolor{deepcerise}{rgb}{0.85, 0.2, 0.53}
\definecolor{denim}{rgb}{0.08, 0.38, 0.74}
\definecolor{crimson}{rgb}{0.86, 0.08, 0.24}
\definecolor{buff}{rgb}{0.94, 0.86, 0.51}
\definecolor{amber(sae/ece)}{rgb}{1.0, 0.49, 0.0}
\definecolor{airforceblue}{rgb}{0.36, 0.54, 0.66}
\definecolor{amethyst}{rgb}{0.6, 0.4, 0.8}
\definecolor{azure(colorwheel)}{rgb}{0.0, 0.5, 1.0}
\definecolor{azure(web)(azuremist)}{rgb}{0.94, 1.0, 1.0}
\definecolor{beige}{rgb}{0.96, 0.96, 0.86}
\definecolor{cornsilk}{rgb}{1.0, 0.97, 0.86}
\definecolor{darkcerulean}{rgb}{0.03, 0.27, 0.49}
\definecolor{babyblue}{rgb}{0.54, 0.81, 0.94}
\definecolor{ballblue}{rgb}{0.13, 0.67, 0.8}
\definecolor{dukeblue}{rgb}{0.0, 0.0, 0.61}
\definecolor{champagne}{rgb}{0.97, 0.91, 0.81}
\definecolor{antiquefuchsia}{rgb}{0.57, 0.36, 0.51}
\definecolor{darkgreen}{rgb}{0, 0.5, 0}

\definecolor{caseblue}{RGB}{95, 125, 150}
\definecolor{casebluebg}{RGB}{238, 243, 247}

\definecolor{cellshade}{HTML}{F0EDFF}

\newcommand{\uf}{\cellcolor[HTML]{FFF9F2}}
\newtcolorbox{casebox}[2][]{
    enhanced,
    breakable,
    colframe=caseblue,
    colback=white,
    coltitle=black,
    fonttitle=\bfseries\small,
    fontupper=\small,
    title={#2},
    attach boxed title to top left={xshift=2mm, yshift*=-3mm},
    boxed title style={
        colback=casebluebg,
        sharp corners,
        frame hidden,
    },
    sharp corners,
    boxrule=0.4mm,
    top=-2.5pt,
    bottom=1.5pt,
    left=2.5pt,
    right=2.5pt,
    width=\linewidth,
    #1
}

\newtcolorbox{casebox2}[2][]{
    enhanced,
    breakable,
    colframe=asparagus,
    colback=white,
    coltitle=black,
    fonttitle=\bfseries\small,
    fontupper=\small,          
    title={#2},
    attach boxed title to top left={xshift=2mm, yshift*=-3mm},
    boxed title style={
        colback=asparagus!10,
        sharp corners,
        frame hidden,
    },
    sharp corners,
    boxrule=0.4mm,
    top=-2.5pt,
    bottom=1.5pt,
    left=2.5pt,
    right=2.5pt,
    width=\linewidth,
    #1
}

\newtcolorbox{darkbluecolorbox}{
  colback=darkcerulean!5,
  colframe=darkcerulean!100,
  coltitle=black,
  fonttitle=\bfseries,
  boxrule=1.5pt,
  arc=3pt,
  left=2pt, right=2pt, top=1pt, bottom=1pt
}

\newcommand{\mstd}[2]{%
#1\raisebox{-0.35ex}{\scriptsize\textcolor{gray}{\,\(\pm #2\)}}%
}

\newcommand{\bestmstd}[2]{%
\textbf{#1}\raisebox{-0.35ex}{\scriptsize\textcolor{gray}{\,\(\pm #2\)}}%
}

\newcommand{\secondmstd}[2]{%
\underline{#1}\raisebox{-0.35ex}{\scriptsize\textcolor{gray}{\,\(\pm #2\)}}%
}

\title{
When Correct Memory Goes Wrong:
Fuzzing Persistent Memory Use in LLM Agents
}

\author{
\textbf{Yuqiao Meng}$^{1}$ \quad
\textbf{Luoxi Tang}$^{1}$ \quad
\textbf{Yingxue Zhang}$^{1}$ \quad
\textbf{Yuchen Yang}$^{2}$ \quad
\textbf{Zhaohan Xi}$^{1}$ \\
\\
$^{1}$Binghamton University, State University of New York,
Binghamton, NY, USA \\
$^{2}$The Pennsylvania State University,
University Park, PA, USA
}

\iclrfinalcopy

\AddToHook{cmd/maketitle/after}{
    \fancyhf{}
    \fancyhead[L]{Preprint.}
    \renewcommand{\headrulewidth}{0.4pt}
    \pagestyle{fancy}
    \thispagestyle{fancy}
}

\begin{document}

\maketitle


\begin{abstract}
Persistent memory helps LLM agents carry information across long interactions, but correct memory can still be used incorrectly when queries change or memory states evolve. Existing work mainly studies memory content errors or evaluates fixed test cases, leaving memory-use failures hard to discover systematically. We formulate this issue as a fuzzing problem and categorize such failures into query-related and memory-state failures. We then develop \system, which starts from memory checkpoints as test seeds, mutates queries or memory states under explicit mutation obligations, validates each mutant, and uses observed memory behavior to guide iterative testing while keeping failure labels outside the search. We evaluate \system across several memory systems against diverse fuzzing
baselines, and further test an output-only setting with API-based LLMs where
memory retrieval is hidden. Across these settings, \system consistently
uncovers more confirmed memory-use failures, showing that its search remains
effective across different memory architectures and even when only final
responses are observable.
\end{abstract}
\section{Introduction}
\label{sec:intro}

Persistent memory is widely used as the external storage of LLM agents to carry information across long and complex interactions, allowing agents to recall stored facts, past experiences, and user context when handling new requests \citep{maharana2024evaluating,ICLR2025_d813d324,hu2026evaluating}. Notably, \textbf{a correct memory can still be used incorrectly}: a small change to a request, such as paraphrasing, may cause the correct memory to be missed or displaced by another record, while a memory update may leave an outdated record ranked above the current one \citep{tang2026trap}. Reliable agentic memory therefore depends on trustworthy stored content and robust use of that content.

However, existing work has mainly studied memory reliability of stored content. Security studies examine how poisoned or injected records enter memory and later affect agent behavior \citep{chen2024agentpoison,dong2026memory}. Diagnostic frameworks further separate retrieval failures from failures in using retrieved memories \citep{yuan2026diagnosing}, and metamorphic testing has begun to test how RAG systems respond when their underlying corpora change \citep{kim2026knowledge}. These studies reveal important memory failures, but they provide limited ways to actively search for cases where an agent uses valid memory incorrectly. Such failures are especially practical because they can arise from ordinary user requests and routine memory updates, even when every stored item is free from poisoning or other failures \citep{xiong2026memory}. As a result, inspections on memory items cannot detect the failure of memory use.

\textbf{This work.}
We highlight that finding memory usage failures is essentially a fuzzing problem, as the same memory system can behave differently across many valid forms of a request and many memory states, while only a small part of this space is covered by test queries \citep{bohme2016coverage,xia2024fuzz4all,kim2026testing}. It is therefore difficult to know in advance which query or memory change will trigger a failure. This motivates a fuzzing-based testing that dynamically generates new cases, observes the resulting memory behavior, and uses that feedback to decide what to test next.

To better support fuzzing design, we first categorize memory-use failures based on what changes during testing. The first group contains \emph{query-triggered} failures, where the memory state is fixed but the request is adapted. For example, two paraphrases asking for the same fact may retrieve different memories, or a request for unsupported information may still retrieve an unrelated record. The second group contains \emph{memory-state} failures, where the request is fixed but the stored memory changes. For example, an update may fail to replace an older value, a conflicting record may be handled incorrectly, or a deleted memory may still be retrieved. This categorization follows the two main factors that determine memory use: what the user asks and what is currently stored in memory. It also gives a simple way to test whether memory use remains correct when either factor changes.

To find these memory-use failures, we develop \system, a fuzzing framework
for persistent agentic memory. \system starts from memory checkpoints built
through normal interactions and constructs test seeds from their associated
queries. It then mutates one input to memory use at a time, covering both
query changes and controlled memory-state changes. Query mutations preserve
the request, redirect it to another supported target, or make it unsupported;
memory-state mutations update, delete, or modify unrelated stored information.
Each mutant is validated before execution, while mutation obligations ensure
that applicable operator-target combinations are exercised systematically.
Observed execution behavior then guides which mutants are retained and
explored in later fuzzing rounds, without using failure labels to direct the
search.

\begin{figure}[t]
    \centering
    \includegraphics[width=0.8\linewidth]{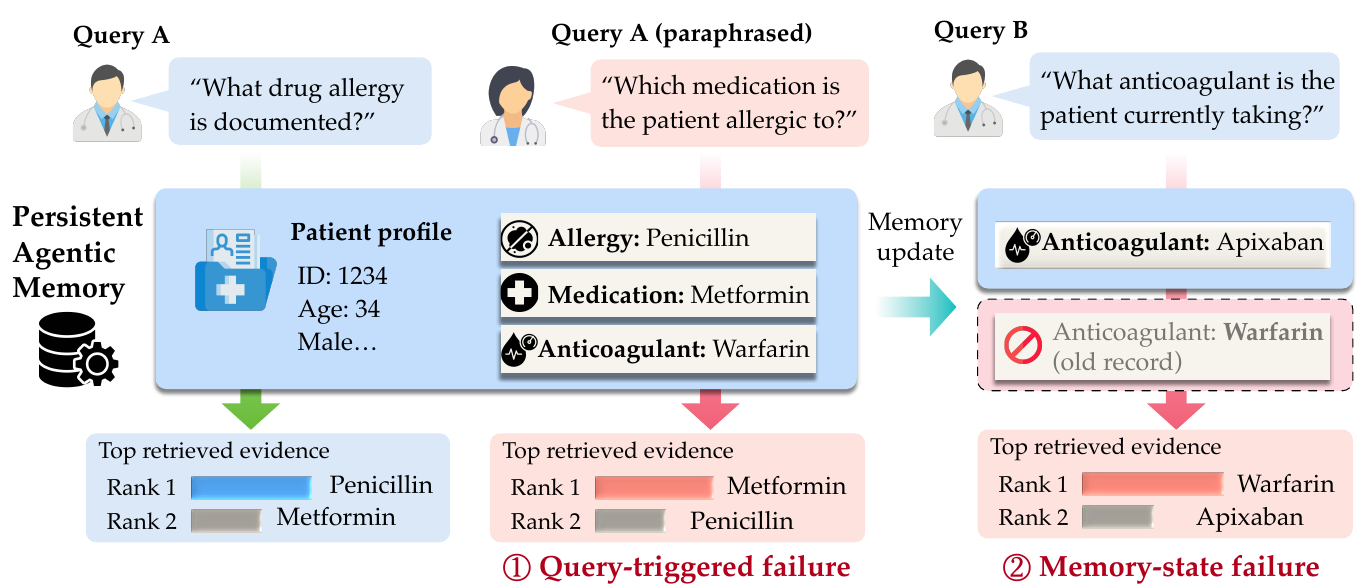}
    \vspace{-5pt}
    \caption{Illustrations of two memory-use failures: (1) paraphrasing a
    query changes the evidence retrieved from the same memory state; (2)
    after a memory update, stale evidence is ranked above the current value.}
    \label{fig:intro}
    \vspace{-5pt}
\end{figure}

We evaluate \system on diverse memory systems
\citep{chhikara2025mem0,xu2026amem,rasmussen2025zep,kang2025memory}
using benchmarks
\citep{maharana2024evaluating,ICLR2025_d813d324}. Our experiments study four
aspects. First, under the same execution budget, \system consistently finds
more unique memory-use failures than diverse baseline methods with different searching mechanisms. The results also
show that broad memory coverage alone is not sufficient: distinct retrieval
behavior within already reached memory remains important for failure
discovery. Second, query and memory-state mutations exhibit different search
dynamics and expose complementary failures; combining them helps \system
continue exploring unique memory-use failures in a long-term execution. Third, ablations
show that the mutation operators contribute differently across memory
backends, while retrieval feedback drives broader exploration or refines the search within explored memory. Finally, we evaluate
an output-only setting in which retrieval is hidden, wherein \system continues to
find more unique and end-to-end failures than baselines,
showing that direct retrieval access improves search guidance and diagnosis
but is not required for memory-use testing.

Overall, we make three contributions:
\begin{itemize}
    \item \textbf{A categorization of memory-use failures.}
    We distinguish query-related and memory-state failures, depending on
    whether incorrect memory use is triggered by a change in the request or
    in the stored memory.

    \item \textbf{A fuzzing framework for persistent memory.}
    We develop \system, which systematically mutates queries and memory
    states, validates each mutant, and uses observed execution behavior to
    guide iterative testing without using failure labels during search.

    \item \textbf{Experimental evaluation.}
    We evaluate \system across multiple memory systems and benchmarks,
    including an output-only setting with hidden retrieval, and show that it
    discovers memory-use failures missed by simpler testing strategies.
\end{itemize}

\section{Related Work}
\label{sec:related}

\textbf{Memory systems and evaluation.}
Persistent memory enables LLM agents to retain, retrieve, update, and organize information across interactions, with designs ranging from episodic memory to structured, dynamically linked, and hierarchical stores~\citep{shinn2023reflexion,chhikara2025mem0,xu2026amem,kang2025memory}. Long-horizon benchmarks evaluate these systems through factual recall, temporal and multi-session reasoning, knowledge updates, abstention, and selective forgetting~\citep{maharana2024evaluating,ICLR2025_d813d324,hu2026evaluating}. Recent empirical work further examines how memory-management operations affect subsequent agent behavior~\citep{xiong2026memory}. Together, these studies substantially broaden memory evaluation, but evaluation remains largely organized around predefined interaction histories and test queries at fixed evaluation points.

\textbf{Reliability and security of agent memory.}
Persistent memory can introduce reliability and security risks when stored information is malicious, stale, conflicting, or otherwise poorly managed \citep{patlan2025context,wei2025memguard,meng2026equimem}. Memory-poisoning and injection studies show that adversarial records can be introduced or activated to influence later agent behavior~\citep{chen2024agentpoison,dong2025minja}, while benchmark and empirical studies expose difficulties involving knowledge updates, forgetting, experience reuse, and error propagation~\citep{ICLR2025_d813d324,hu2026evaluating,xiong2026memory}. Existing work therefore establishes that memory reliability depends on both stored content and its subsequent use, but primarily studies predefined attacks, benchmark cases, or controlled memory-management operations.

\textbf{Fuzzing and testing of LLM-based systems.}
LLMs have been incorporated into fuzzing both as generators of test inputs for software systems and as systems under test through prompt mutation~\citep{deng2023titanfuzz,xia2024fuzz4all,yu2024llmfuzzer,shao2026promptfuzz}. Retrieval-aware testing has further introduced coverage criteria and metamorphic mutations to explore RAG behavior beyond fixed queries~\citep{kim2026testing,kim2026knowledge}. These studies demonstrate the value of coverage-guided and mutation-based testing, but primarily target software inputs, prompt behavior, or document-level retrieval systems rather than persistent agent memory in which both queries and accumulated memory states evolve across interactions.
\section{Categorization of Memory-Use Failures}
\label{sec:problem}

Persistent memory affects an agent through the information that is selected from memory and used to handle a new request. We use
\(
U(q,M)
\)
to denote the observable memory-use behavior of an agent when processing query $q$ with memory state $M$. Depending on the system, $U(q,M)$ may
include the memories retrieved for the query, their ranking, or the
memory-dependent response produced by the agent. We do not assume that a
memory-use failure is caused by any particular internal component.

We categorize memory-use failures by changing one of the two inputs to
$U(q,M)$ while keeping the other fixed. This gives two categories:

\textbf{I. Query-related failures} occur when the memory state remains fixed,
but memory use does not correctly follow an adaptation from $q$ to
$q'$. The expected behavior depends on the relation between the two
queries. For example, if $q'$ is a paraphrase of $q$, the same relevant
memory should continue to support the request; if $q'$ asks for unsupported
information, the agent should not rely on unrelated memory. We write this
requirement as
\fbox{\(
\mathcal{C}_Q\!\left(U(q,M), U(q',M); q,q'\right)
\)},
where $\mathcal{C}_Q$ specifies the relation expected from the query
change, such as preserving memory use under equivalent queries or avoiding
unsupported memory use for an unanswerable query. A query-related failure
occurs when this condition is violated.

\textbf{II. Memory-state failures} occur when the query remains fixed, but
memory use does not correctly follow a controlled change from $M$ to
$M'$. A relevant update, deletion, or addition should be reflected when
the agent later uses memory, while an unrelated change should not disturb
the information used for the same request. We write this requirement as
\fbox{\(
\mathcal{C}_M\!\left(U(q,M), U(q,M'); M,M'\right)
\)},
where $\mathcal{C}_M$ specifies the relation expected from the memory
change, such as preferring the updated value after a replacement or no
longer using an entry after it has been deleted. A memory-state failure
occurs when this condition is violated.

This categorization is based on the change that triggers the failure, thus agnostic to how the memory system is integrated to LLM agents. It therefore applies
to systems with explicit retrieval interfaces as well as systems whose
memory use mechanism is hidden. The mutation operators in
Section~\ref{sec:method} instantiate these two categories with concrete
query and memory changes.
\section{Method}
\label{sec:method}

\begin{figure}[t]
    \centering
    \includegraphics[width=0.98\linewidth]{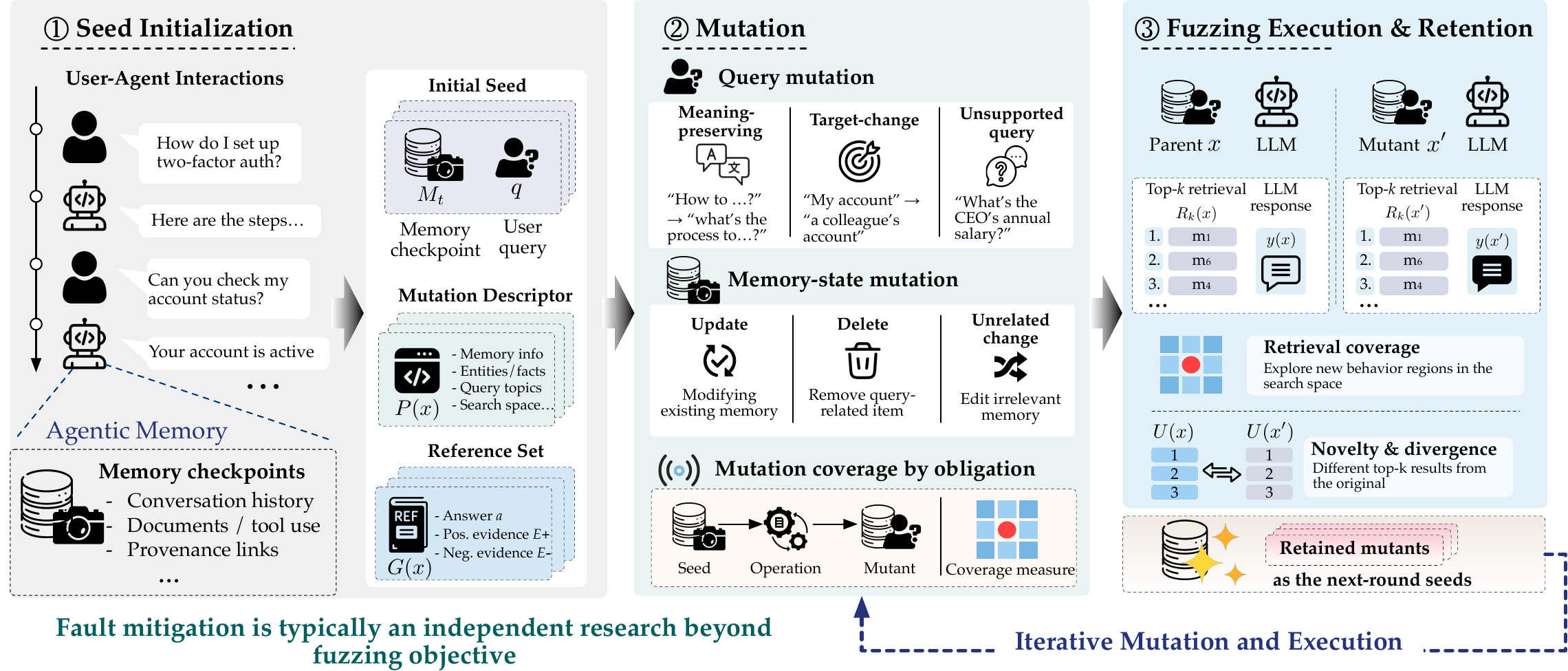}
    \vspace{-7pt}
    \caption{Overview of \system. \system builds seeds and references from
    memory checkpoints, generates validated query and memory-state mutations,
    and executes each parent-mutant pair to detect memory-use failures.
    Retrieval coverage and divergence guide later fuzzing rounds,
    while failure labels remain outside the fuzzing search.}
    \label{fig:method}
    \vspace{-10pt}
\end{figure}

We develop \system, a fuzzing framework for discovering the memory-use
failures defined in \S\ref{sec:problem}. While iterative mutation and
execution are standard in fuzzing \citep{xia2024fuzz4all,deng2023titanfuzz,shao2026promptfuzz}, \system makes persistent memory use as
the fuzzing target by treating both the query and evolving memory state as
first-class inputs. As outlined in
Figure~\ref{fig:method}, \system first constructs seeds from memory checkpoints
(\S\ref{subsec:seed-init}). It then generates and validates query or
memory-state mutations under an explicit mutation-coverage requirement
(\S\ref{subsec:mutation}). Finally, it executes each valid parent--mutant
pair, checks whether the mutation exposes a memory-use failure, and retains
useful mutants for later rounds (\S\ref{subsec:execution}).

\subsection{Seed Initialization}
\label{subsec:seed-init}

\begin{wrapfigure}{r}{0.43\columnwidth}
\vspace{-1.2em}
\begin{minipage}{\linewidth}
\scriptsize
\begin{algorithm}[H]
  \SetAlgoLined
  \caption{Seed Construction
  \label{algo:seed}}

  \KwIn{Task history $H$; checkpoint times $\mathcal{T}$;
        query sets $\{\mathcal{Q}_t\}$; provenance $Z$}
  \KwOut{Seeds $\mathcal{S}_0$; descriptors $P$; references $G$}

  $\mathcal{S}_0\leftarrow\emptyset$\;

  \ForEach{$t\in\mathcal{T}$}{
    $M_t\leftarrow\textsc{Checkpoint}(H_{1:t})$\;
    \ForEach{$q\in\mathcal{Q}_t$}{
      $x\leftarrow(M_t,q)$\;
      $P(x)\leftarrow
      \textsc{BuildDescriptor}(q,H_{1:t},Z)$\;
      $G(x)\leftarrow
      \textsc{BuildReference}(q,H_{1:t},Z)$\;
      $\mathcal{S}_0\leftarrow\mathcal{S}_0\cup\{x\}$\;
    }
  }

  \Return $\mathcal{S}_0,P,G$\;
\end{algorithm}
\end{minipage}
\vspace{-1.0em}
\end{wrapfigure}

\system starts from checkpoints taken after an agent has accumulated memory
through normal interactions. Let
\(H_{1:t}=\langle h_1,\ldots,h_t\rangle\) denote the interaction history up
to time \(t\), and let \(M_t\) denote the resulting memory state. If the
system supports state copying, \system creates isolated copies of \(M_t\).
Otherwise, it recreates the same state in a separate session by replaying
\(H_{1:t}\). The original task and its memory are left unchanged.

A seed is \(x=(M_t,q)\), where \(q\) is a query whose answer may depend on
information in \(M_t\). As shown in Algorithm~\ref{algo:seed}, each seed has
two separate records. The mutation descriptor \(P(x)\) contains only
information needed to construct valid mutations, such as query slots,
available replacement targets, absent targets, and supported memory
operations. \(P(x)\) is used during mutation
(\S\ref{subsec:mutation}) and contains no expected answer. Each seed also
has a reference set \(G(x)=(a,E^{+},E^{-})\) for evaluation
(\S\ref{subsec:execution}). Here, \(a\) is the expected answer,
\(E^{+}\) contains valid evidence from memory, and \(E^{-}\) contains
evidence made invalid by a memory operation, such as an update or deletion.
By default, \(E^{-}=\varnothing\). We use \(a=\bot\) when the memory does
not contain information that supports an answer to \(q\).


\subsection{Mutation}
\label{subsec:mutation}

Given a seed \(x=(M_t,q)\), \system changes one input to memory use at a
time. A \textbf{query mutation} produces \(x'=(M_t,q')\), while a
\textbf{memory-state mutation} produces \(x'=(M_t',q)\). We change only one
input in each pair so that an observed difference can be attributed to the
query change or the memory-state change. A retained mutant may become a
seed in a later round, so different changes can accumulate across rounds.

\textbf{I. Query mutation:}
A query mutation keeps \(M_t\) fixed. \system first decides how \(q'\)
should differ from \(q\), then constructs a query that satisfies this
change. \textbf{(I-1) Meaning-preserving mutation.} The requested entity,
attribute, time scope, answerability, and expected answer remain unchanged,
so \(a'=a\). Operators include paraphrasing, shortening removable context,
adding nonrestrictive context, and changing syntactic form.
\textbf{(I-2) Target-changing mutation.} The mutant changes one requested
target while preserving the rest of the query. If \(z\) is a replaceable
entity, attribute, or time scope and \(z'\) is another target listed in
\(P(x)\), then \(q'=q[z\leftarrow z']\). We construct \(G(x')\) from the
provenance of \(z'\), while the search does not receive this reference.
\textbf{(I-3) Unsupported mutation.} The mutant replaces the requested
target with \(z_{\emptyset}\), whose absence is recorded in \(P(x)\), and
therefore \(a'=\bot\). To avoid confusing missing memory with parametric
knowledge of the base LLM, we use only \emph{memory-exclusive} unsupported
targets: synthetic or session-local entities and attributes whose answers
are not available from public model knowledge. Their absence is established
before execution and is never inferred from a failed retrieval or response.

\textbf{II. Memory-state mutation:}
A memory-state mutation keeps \(q\) fixed and applies a controlled operation
\(\delta\) to an isolated copy of \(M_t\), producing
\(M_t'=\delta(M_t)\). If the system exposes memory operations, \system uses
them directly. If memory changes only through interaction, the required
interaction is appended to the history that produced \(M_t\), and \(M_t'\)
is created in a separate session. Specifically, we design three operations:
\textbf{(II-1) Update} adds a new value \(v_{\mathrm{new}}\) for an existing
entity-attribute pair. For
\(v_{\mathrm{old}}\rightarrow v_{\mathrm{new}}\), we set
\(a'=v_{\mathrm{new}}\) and place the old evidence in \(E'^{-}\).
\textbf{(II-2) Deletion} removes an existing fact; the deleted evidence is
placed in \(E'^{-}\), and \(a'=\bot\) if no valid support remains.
\textbf{(II-3) Unrelated change} modifies information outside the target of
\(q\), so \(G(x')=G(x)\). Its expected effect is no change in memory use for
the fixed query. For mutation types that do not create explicitly invalid
evidence, \(E'^{-}=\varnothing\).

Detailed construction and validation methods are given in
Appendix~\ref{app:mutation-implementation}. Generated mutants are accepted only when the
operator-specific conditions can be checked independently of the mutant's
retrieval result or final response.

\textbf{Mutation coverage:}
The space of natural-language queries is unbounded, so \system defines
coverage over the finite mutation opportunities available from the current
seed corpus. At the beginning of round \(r\), the seed corpus
\(\mathcal{S}_r\) is fixed. We define a mutation obligation
\(\omega=(x,o,u)\), which pairs a seed \(x\), an applicable operator \(o\),
and one concrete mutation target \(u\). For a query mutation, \(u\) is the
query element or replacement target being changed; for a memory-state
mutation, \(u\) identifies the memory entry involved in the operation. Let
\(\mathcal{U}_o(x)\) denote the valid targets of operator \(o\) for seed
\(x\). We have:
\vspace{-8pt}
\begin{equation}
\Omega_r =
\left\{
(x,o,u)
\;\middle|\;
x\in\mathcal{S}_r,\;
u\in\mathcal{U}_o(x),\;
\mathrm{pre}_o(x,u)=1
\right\}
\quad
\mathrm{Cov}_{\mathrm{mut}}^{(r)}
=
\frac{|\Omega_r^{\mathrm{exec}}|}{|\Omega_r|}
\label{eq:mutation-coverage}
\end{equation}
where \(\mathrm{pre}_o(x,u)\) specifies whether operator \(o\) can be
applied to seed \(x\) at target \(u\), and
\(\Omega_r^{\mathrm{exec}}\) contains obligations that have produced at
least one validated and executed mutant.

Each uncovered obligation remains in the generation queue until it produces
a valid mutant or reaches a fixed generation-attempt limit \(A\). Extra
variants are generated only after the uncovered obligations have been tried.
Thus, if every obligation yields a valid mutant within \(A\) attempts and
the round has enough execution budget for one such mutant per obligation,
then \(\mathrm{Cov}_{\mathrm{mut}}^{(r)}=1\). This guarantee applies only
to the operators and targets represented in \(\Omega_r\). Retained mutants
are added to \(\mathcal{S}_{r+1}\), where they may create new obligations.

\subsection{Fuzzing Execution and Evaluation}
\label{subsec:execution}

For each validated pair \((x,x')\), both executions start from independent
copies of the parent checkpoint \(M_t\). For a query mutation, \(M_t\) is
copied twice and the two queries are evaluated separately. For a
memory-state mutation, one copy remains at \(M_t\), while \(\delta\) is
applied to the other to produce \(M_t'\). If copying is unavailable, the
corresponding checkpoint is recreated in separate sessions. This prevents
one test query, including any memory write it may trigger, from changing the
other execution.

\begin{wrapfigure}{r}{0.43\columnwidth}
\vspace{-14pt}
\begin{minipage}{\linewidth}
\scriptsize
\begin{algorithm}[H]
  \SetAlgoLined
  \caption{Execute and Retain
  \label{algo:execute}}

  \KwIn{Parent $x$; validated mutant $x'$; operator $o$}
  \KwOut{Failure result $F_o$; retain flag $b$}

  $U(x),U(x')\leftarrow
  \textsc{IsolatedExec}(x,x')$\;

  \tcp{Evaluation}
  $F_o\leftarrow
  \textsc{CheckFailure}
  (o,U(x),U(x'),G(x),G(x'))$\;

  \tcp{Fuzzing feedback}
  \eIf{retrieval is visible}{
    $b\leftarrow
    \textsc{RetrievalRetention}(x,x',o)$\;
  }{
    $b\leftarrow
    \textsc{ResponseRetention}(x')$\;
  }

  Store $F_o$ separately\;

  \If{$b=1$}{
    Add $x'$ to next-round seeds\;
  }

  \Return $F_o,b$\;
\end{algorithm}
\end{minipage}
\vspace{-1.0em}
\end{wrapfigure}
Let \(U(x)\) denote the observed memory-use behavior. For a
retrieval-visible system, we record
\(U(x)=(R_k(x),y(x))\), where
\(R_k(x)=[m_1,\ldots,m_k]\) is the ranked top-\(k\) retrieved memory and
\(y(x)\) is the final response. For a retrieval-hidden system, only
\(y(x)\) is available.

\textbf{Evaluation of Memory-Use Failure:}
For retrieval-visible systems, entries in \(R_k(x)\) are matched with
\(E^{+}\) and \(E^{-}\) using memory identifiers or provenance. We define
\(r^{+}(x)=\min_{m\in E^{+}}\mathrm{rank}(m,R_k(x))\), and
\(r^{-}(x)=\min_{m\in E^{-}}\mathrm{rank}(m,R_k(x))\) when
\(E^{-}\neq\varnothing\); otherwise \(r^{-}(x)=\infty\). A missing item
also has rank \(\infty\). In regular agent use, we check \(\mathrm{Ans}(y(x),a)\) to determine whether
the final response gives the expected answer. In fuzzing, for each mutation operator \(o\), a memory-use
failure occurs when the observed behavior violates an expectation \(\mathcal{C}_o\) between the parent and mutant:
\vspace{-4pt}
\begin{equation}
F_o(x,x') =
\mathbf{1}\!\left[
\neg\mathcal{C}_o
\left(
U(x),U(x');
G(x),G(x')
\right)
\right]
\label{eq:failure-detection}
\end{equation}
where \(F_o=1\) denotes a memory-use failure. For query mutations,
\(F_o=1\) if \textbf{(I-1)} a meaning-preserving mutant no longer preserves
the expected answer or valid support; \textbf{(I-2)} a target-changing
mutant does not follow \(G(x')\); or \textbf{(I-3)} an unsupported query
receives a factual answer despite \(a'=\bot\). For memory-state mutations,
\(F_o=1\) if \textbf{(II-1)} updated evidence is missing
(\(r^{+}(x')=\infty\)) or stale evidence is ranked above it
(\(r^{-}(x')<r^{+}(x')\)); \textbf{(II-2)} deleted evidence in
\(E'^{-}\) is still retrieved or used; or \textbf{(II-3)} the answer or
valid support changes even though the query-relevant memory remains
unchanged.

When retrieval is visible, \system can further localize where the failure
appears. Missing valid evidence or invalid evidence ranked above it indicates
a failure at memory selection. If valid evidence is retrieved but the final
response is incorrect, the failure appears after selection. \system does not
attribute such failures to a specific internal LLM operation, since follow-up
diagnosis is outside the fuzzing objective. For systems that hide retrieval
details, \system reports only the end-to-end memory-use failure.

\textbf{Seed retention and iteration:}
Failure detection is kept separate from fuzzing: \(F_o\) is stored for
evaluation but never used to retain or prioritize seeds. For systems with
visible retrieval, \system uses three feedback indicators: \emph{coverage} for
reaching unseen memory entries, \emph{novelty} for new top-\(k\) retrieval
patterns, and \emph{divergence} for large parent-mutant ranking changes in
meaning-preserving mutations. Coverage and novelty determine retention, while
divergence further guides prioritization. For API-based systems with hidden
retrieval, \system instead retains mutants that produce previously unseen
response behavior. At the end of round \(r\), retained mutants are added to
\(\mathcal{S}_{r+1}\) and used to generate new mutation obligations. More details are given in Appendix~\ref{app:seed-retention}.
\section{Experiment}
\label{sec:experiments}

\textbf{Memory systems:}
We evaluate on four retrieval-visible memory systems:
Mem0~\citep{chhikara2025mem0}, A-Mem~\citep{xu2026amem},
Graphiti~\citep{rasmussen2025zep}, and MemOS~\citep{kang2025memory}.
These systems expose the memories selected for each query and form the
default setting. We also separately removes retrieval visibility
and evaluates \system using only the final response (\S\ref{subsec:rq4}).

\textbf{Benchmarks:}
We use LoCoMo~\citep{maharana2024evaluating} and
LongMemEval-S~\citep{ICLR2025_d813d324}, which provide long interaction
histories together with memory-dependent queries and references. We construct
memory checkpoints and initial seeds from the benchmark interactions following
\S\ref{subsec:seed-init}.

\textbf{Baselines:}
We compare \system with four baseline strategies: \textit{(i) Random Mutation} defines a uniform obligations across mutants. \textit{(ii) Unguided LLM} uses the LLM itself to generate and select mutants. \emph{(iii) LLM-as-Judge} ranks valid mutants by their estimated likelihood of exposing a failure, without access to references. \emph{(iv) Coverage-Guided} retains and prioritizes
mutants using newly reached memory entries only. Other settings (seed, budget) are same across baselines and \system.

\textbf{Metrics.}
We evaluate fuzzing under an execution budget \(B\) by two complementary metrics: 
\emph{(i) UF@\(B\)} counts the unique memory-use failures found within
\(B\) valid executions. \emph{(ii) Cov@\(B\)} measures the fraction of memory entries retrieved at least once. 



\subsection{RQ1: How effective is \system at memory-use failure discovery?}
\label{subsec:rq1}

We first compare overall fuzzing effectiveness under a fixed execution
budget. Table \ref{tab:rq1} shows results across all fuzzing methods on both benchmarks and four memory systems, wherein we also consider two adaptations of \system: \system-Q and \system-M uses query-only or memory-state-only mutations, respectively. We have the following observations and insights:

\begin{table}[t]
\centering
\scriptsize
\caption{
Performance under a fixed execution budget \(B\)
(\(B{=}8000\) for LoCoMo and \(B{=}4000\) for LongMemEval-S).
Results report mean and standard deviation over three runs. Appendix \ref{app:mutation-cases} also provides some case studies of found memory-use failures.
}
\label{tab:rq1}
\renewcommand{\arraystretch}{1.05}
\setlength{\tabcolsep}{4pt}
\resizebox{\textwidth}{!}{
\begin{tabular}{ll|cccccccc}
\toprule
& \textbf{Method}
& \multicolumn{2}{c}{\textbf{Mem0}} & \multicolumn{2}{c}{\textbf{A-Mem}} & \multicolumn{2}{c}{\textbf{Graphiti}} & \multicolumn{2}{c}{\textbf{MemOS}} \\
\cmidrule(lr){3-4}\cmidrule(lr){5-6}\cmidrule(lr){7-8}\cmidrule(lr){9-10}
& & \textbf{UF@\(B\)} & \textbf{Cov@\(B\)} & \textbf{UF@\(B\)} & \textbf{Cov@\(B\)} & \textbf{UF@\(B\)} & \textbf{Cov@\(B\)} & \textbf{UF@\(B\)} & \textbf{Cov@\(B\)} \\
\midrule
\multirow{7}{1.2em}{\rotatebox[origin=c]{90}{\textbf{LoCoMo}}}
& Random Mutation & \mstd{12.7}{1.5} & \mstd{0.40}{.02} & \mstd{21.7}{5.1} & \mstd{0.24}{.08} & \mstd{18.7}{3.2} & \mstd{0.27}{.12} & \mstd{15.0}{4.6} & \mstd{0.34}{.10} \\
& Unguided LLM & \mstd{22.0}{2.0} & \mstd{0.47}{.03} & \mstd{29.7}{3.8} & \mstd{0.39}{.08} & \mstd{27.3}{4.0} & \mstd{0.35}{.10} & \mstd{26.0}{3.0} & \mstd{0.45}{.12} \\
& LLM-as-Judge & \mstd{23.3}{2.5} & \mstd{0.41}{.01} & \mstd{35.7}{2.3} & \mstd{0.38}{.04} & \mstd{23.3}{1.5} & \mstd{0.34}{.02} & \mstd{25.3}{0.6} & \mstd{0.45}{.02} \\
& Coverage-Guided & \mstd{27.7}{4.0} & \secondmstd{0.59}{.03} & \mstd{31.3}{2.5} & \secondmstd{0.54}{.05} & \mstd{30.3}{3.5} & \bestmstd{0.47}{.04} & \mstd{30.0}{1.0} & \secondmstd{0.65}{.02} \\
& \uf U-Fuzz-Q & \uf\secondmstd{33.0}{2.6} & \uf\mstd{0.57}{.07} & \uf\secondmstd{57.0}{1.0} & \uf\mstd{0.51}{.03} & \uf\mstd{39.7}{2.9} & \uf\mstd{0.42}{.05} & \uf\secondmstd{41.3}{2.5} & \uf\mstd{0.57}{.03} \\
& \uf U-Fuzz-M & \uf\mstd{27.0}{2.0} & \uf\mstd{0.54}{.03} & \uf\mstd{48.3}{2.1} & \uf\mstd{0.45}{.06} & \uf\secondmstd{51.3}{2.1} & \uf\mstd{0.42}{.04} & \uf\mstd{27.3}{0.6} & \uf\mstd{0.54}{.01} \\
& \uf \textbf{U-Fuzz} & \uf\bestmstd{45.3}{1.2} & \uf\bestmstd{0.62}{.02} & \uf\bestmstd{73.3}{3.1} & \uf\bestmstd{0.55}{.02} & \uf\bestmstd{62.0}{2.0} & \uf\secondmstd{0.46}{.03} & \uf\bestmstd{53.7}{1.2} & \uf\bestmstd{0.67}{.02} \\
\midrule
\multirow{7}{1.2em}{\rotatebox[origin=c]{90}{\textbf{LongMemEval-S}}}
& Random Mutation & \mstd{9.3}{4.9} & \mstd{0.29}{.07} & \mstd{16.0}{3.5} & \mstd{0.34}{.12} & \mstd{17.3}{4.5} & \mstd{0.25}{.09} & \mstd{12.7}{5.9} & \mstd{0.33}{.08} \\
& Unguided LLM & \mstd{26.0}{4.4} & \mstd{0.42}{.06} & \mstd{32.0}{3.6} & \mstd{0.42}{.05} & \mstd{23.7}{4.6} & \mstd{0.34}{.08} & \mstd{25.7}{4.2} & \mstd{0.44}{.07} \\
& LLM-as-Judge & \mstd{21.7}{1.5} & \mstd{0.38}{.04} & \mstd{39.3}{1.2} & \mstd{0.39}{.04} & \mstd{32.0}{3.0} & \mstd{0.32}{.03} & \mstd{23.0}{2.6} & \mstd{0.40}{.03} \\
& Coverage-Guided & \mstd{25.0}{3.5} & \secondmstd{0.60}{.02} & \mstd{33.3}{3.2} & \secondmstd{0.55}{.03} & \mstd{30.7}{3.5} & \secondmstd{0.46}{.05} & \mstd{29.0}{2.6} & \secondmstd{0.62}{.02} \\
& \uf U-Fuzz-Q & \uf\secondmstd{31.3}{2.5} & \uf\mstd{0.53}{.03} & \uf\secondmstd{48.0}{2.0} & \uf\mstd{0.53}{.04} & \uf\mstd{36.7}{1.5} & \uf\mstd{0.40}{.02} & \uf\secondmstd{35.7}{1.2} & \uf\mstd{0.57}{.02} \\
& \uf U-Fuzz-M & \uf\mstd{28.7}{2.5} & \uf\mstd{0.53}{.03} & \uf\mstd{45.3}{3.2} & \uf\mstd{0.48}{.03} & \uf\secondmstd{42.7}{2.9} & \uf\mstd{0.42}{.01} & \uf\mstd{26.0}{3.0} & \uf\mstd{0.53}{.01} \\
& \uf \textbf{U-Fuzz} & \uf\bestmstd{40.3}{2.5} & \uf\bestmstd{0.62}{.02} & \uf\bestmstd{66.3}{0.6} & \uf\bestmstd{0.57}{.02} & \uf\bestmstd{53.0}{1.0} & \uf\bestmstd{0.48}{.00} & \uf\bestmstd{42.7}{1.2} & \uf\bestmstd{0.66}{.03} \\
\bottomrule
\end{tabular}}
\end{table}

\textbf{Behavior-guided fuzzing improves failure discovery without reducing
the search to coverage alone.}
Table~\ref{tab:rq1} shows that \system consistently discovers more unique
memory-use failures than the baseline strategies across benchmarks and
memory systems. Coverage-Guided is effective at expanding the set of reached
memories, but this does not translate directly into the same level of
failure discovery. \system  maintains comparable or stronger
memory coverage while also exploring different behaviors around the memories
already reached. This distinction matters because fuzzing methods that reach an entry only
ensures that memory is accessible; it does not inspect how reliably the agent
uses that entry under different queries and memory states.

\textbf{Query and memory-state mutations are complementary.}
The variants of \system show complementary behavior. Query mutations are
more effective on Mem0, A-Mem, and MemOS, because retrieval in these
systems is more directly shaped by query formulation. Memory-state mutations
are stronger on Graphiti, where updates and deletions can alter the
graph structure and subsequent retrieval paths.
This supports our categorization in \S\ref{sec:problem}: memory-use
failures can be triggered by either query changes or memory-state changes,
with their relative importance depending on the backend.

\textbf{Execution feedback is necessary.}
LLM-as-Judge is competitive in some settings, but its effectiveness varies
across memory systems and benchmarks. This suggests that asking an LLM to predict failure-triggering mutants is not always reliable. In contrast, feedback from the system’s actual execution behavior provides a firmer basis for guiding later fuzzing rounds.

\subsection{RQ2: How does fuzzing effectiveness evolve with the execution budget?}
\label{subsec:rq2}

\begin{figure}[t]
    \centering
    \includegraphics[width=\linewidth]{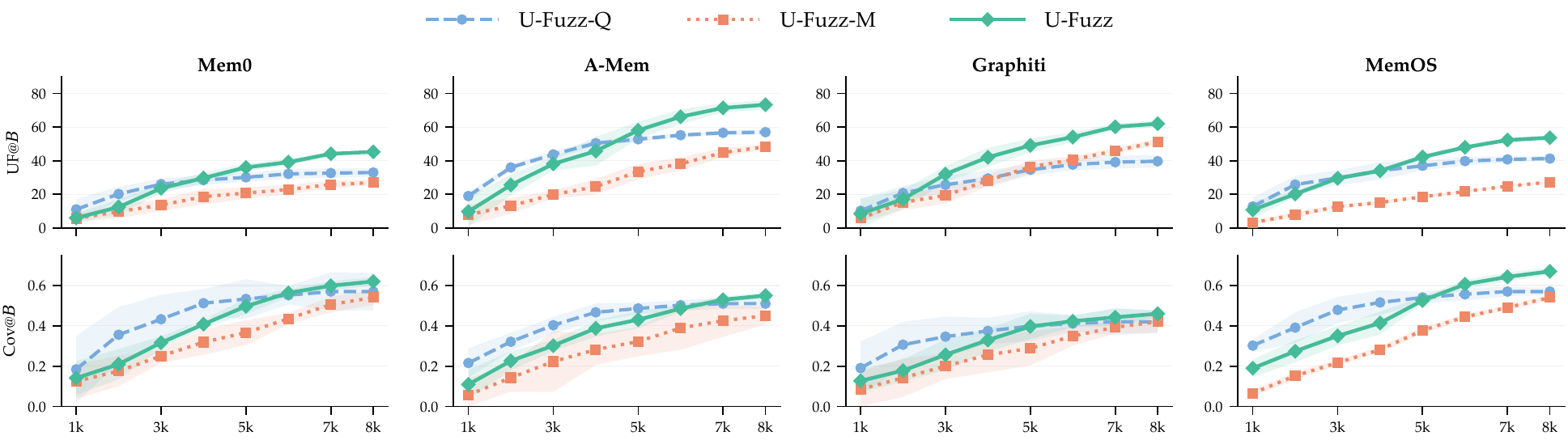}
    \vspace{-5pt}
    \caption{
    Fuzzing progress as the execution budget (x-axis) increases on LoCoMo.
    Each column corresponds to one memory system. Shaded regions indicate
    one standard deviation over three runs.
    }
    \label{fig:rq2_budget}
\end{figure}

We next study how \system's performance changes with the execution budget. Figure~\ref{fig:rq2_budget} shows the results across memory systems.

\textbf{The two mutation classes show different dynamics.}
Query-only fuzzing (\system-Q) progresses faster at the beginning because it can reuse an
existing checkpoint and immediately inspect different query formats. However, it tends to saturate earlier as the available query variations become exhausted. Memory-state fuzzing (\system-M) starts more slowly because each
mutant requires constructing and validating a new memory state, but it
continues to uncover failures that \system-Q cannot reach. The two
variants therefore complement each other in longitudinal fuzzing runs.

\textbf{\system remains effective as either search direction starts to saturate.}
The full \system combines both mutation classes, so later fuzzing rounds can
continue through memory-state changes when query exploration slows, and vice
versa. This makes the search more comprehensive over longer budgets and helps
avoid early saturation around a single type of memory-use behavior.

\subsection{RQ3: How Does Each Component of \system Influence to Its Effectiveness?}
\label{subsec:rq3}

We ablate \system to understand
how each individual component influence to it's performance. On the mutation side, we independently remove
meaning-preserving (MP), target-changing (TC), unsupported (UNS), update
(UPD), deletion (DEL), or unrelated-change (URC) mutations. On the execution side, we independently remove coverage (Cov), retrieval novelty (Nov), or
parent-mutant divergence (Div).

\begin{figure}[t]
    \centering
    \includegraphics[width=0.97\linewidth]{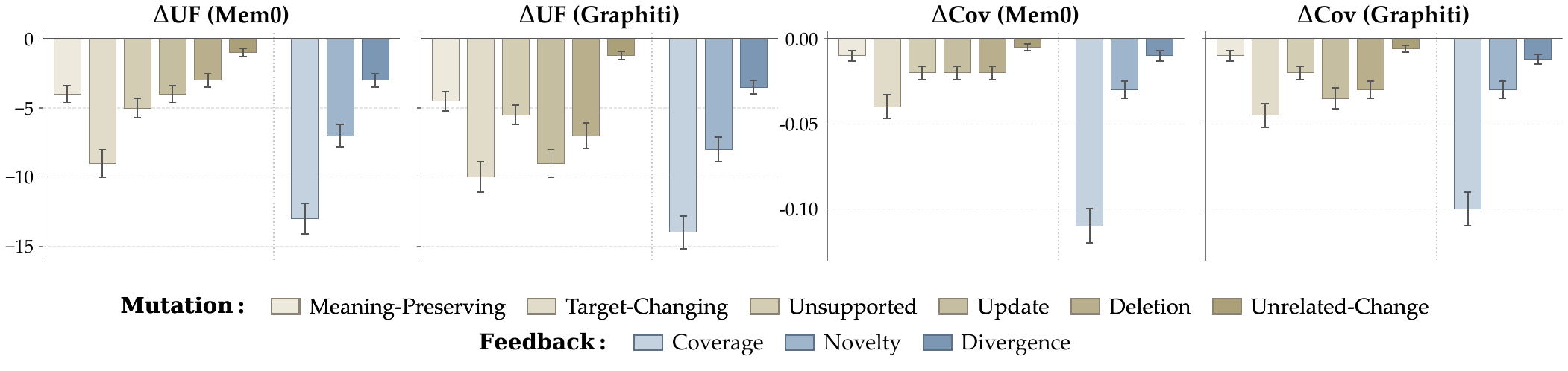}
    \vspace{-3pt}
    \caption{
    Component ablation on LoCoMo.
    Each bar reports the change in UF@\(B\) or Cov@\(B\)  after removing one component from the full \system.
    \emph{Mutation} ablations remove one query- or memory-state operator,
    while \emph{Feedback} ablations remove one search-feedback indicator.
    Error bars show one standard deviation over three runs.
    }
    \label{fig:rq3-ablation}
\end{figure}

\textbf{Different mutation operators expose different failure cases.}
Figure~\ref{fig:rq3-ablation} shows that removing any mutation operator
reduces failure discovery, though the impact varies across backends.
Target-changing contributes most among query mutations because it redirects
the request to other supported targets and expands the memory being tested.
Meaning-preserving and unsupported mutations affect coverage less, but probe
failure types that target changes do not capture. Memory-state mutations show
stronger backend dependence: update and deletion matter more on Graphiti than
on Mem0, suggesting that the two systems respond differently to evolving
memory states. Unrelated change has a smaller effect, as expected from its
role as a non-interference test rather than a mechanism for reaching new
relevant memory.

\textbf{Execution feedback serves different roles.}
Coverage helps \system reach memory entries that have not been tested before.
Retrieval novelty captures new top-\(k\) retrieval patterns among entries that
may already be covered, while parent--mutant divergence measures how much the
ranking changes after a meaning-preserving mutation. The ablation results show
that coverage is the main driver of broader exploration, whereas novelty and
divergence help \system continue finding distinct behaviors within the
explored memory space.

\subsection{RQ4: Does \system remain effective when retrieval is hidden?}
\label{subsec:rq4}

\begin{wrapfigure}{r}{0.35\linewidth}
    \centering
    \vspace{-15pt}
    \includegraphics[width=\linewidth]{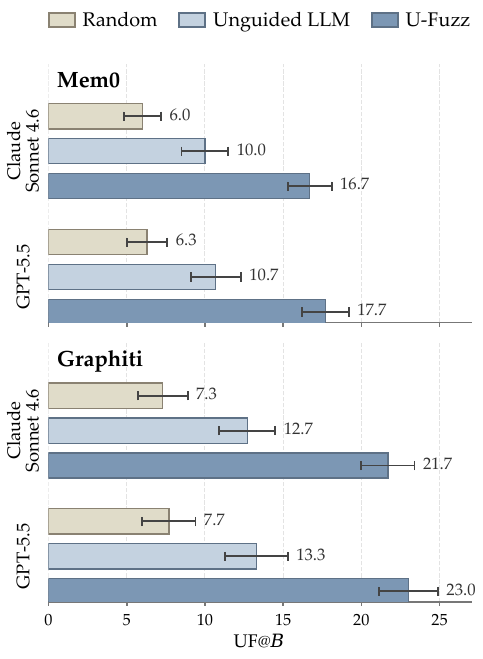}
    \vspace{-18pt}
    \caption{
    Failure discovery under output-only observability on LoCoMo. Each bar reports UF@\(B\)  \(B=2000\), wherein we prompt LLMs to output reasoning and responses. Error bars indicate one standard deviation over three runs.
    }
    \label{fig:rq4-hidden}
    \vspace{-5pt}
\end{wrapfigure}
The previous experiments assume access to ranked retrieval results. We finally
consider an output-only setting in which the memory system executes normally,
but its retrieved entries and rankings are hidden from the fuzzer. Thus,
\system observes \(U(x)=y(x)\) rather than \(U(x)=(R_k(x),y(x))\).
We place Mem0 and Graphiti behind this interface and evaluate them on LoCoMo
with GPT-5.5 and Claude Sonnet 4.6 as the LLM backends. \system retains the
same mutation-obligation schedule, but replaces retrieval-based retention with
novel response behavior as described in \S\ref{subsec:execution}. Each
configuration uses \(B=2000\) and compares Random Mutation, Unguided LLM, and
the output-only version of \system. Coverage-Guided is omitted because memory
coverage is not observable. Failure confirmation still uses \(G(x)\),
\(G(x')\), and the mutation-specific expectation \(\mathcal{C}_o\), but any
failure can only be reported end to end.

\textbf{\system remains effective with only final responses as feedback.}
Figure~\ref{fig:rq4-hidden} shows a consistent advantage for \system across
both memory systems and LLM backends, even after retrieval-level feedback is
removed. This suggests that the search does not depend entirely on observing
which memory entries are retrieved. Mutation obligations still ensure that
different query and memory-state changes are exercised systematically, while
response novelty provides a weaker but still useful signal for deciding which
mutants should continue to later rounds.

\textbf{Hidden retrieval mainly reduces diagnosis process, not the testability.}
The output-only setting still allows \system to confirm whether a mutation
inspects failures of memory-use behavior from the final responses. What it loses
is the ability to explain where that failure occurs. Without \(R_k\), the same
incorrect response may result from retrieving the wrong memory or from
misusing the right one. The results therefore suggest that direct retrieval
access is valuable for finer search guidance and localization, but is not a
requirement for discovering end-to-end memory-use failures.
\section{Conclusion}
\label{sec:conclusion}

We study a reliability problem in persistent-memory agents that is easy to
overlook: even when stored memory is correct, an agent may still use it
incorrectly as queries or memory states change. We formulate this problem as
fuzzing and introduce \system, which systematically mutates both queries and
memory states, validates each test case, and uses observed execution behavior
to guide iterative exploration without relying on failure labels during
search. Across multiple memory systems and benchmarks, \system uncovers
memory-use failures that simpler testing strategies miss, and remains useful
even when retrieval is hidden. More broadly, our results suggest that reliable
agent memory should be evaluated not only by what is stored, but also by how
that memory is used under changing requests and evolving states.

\bibliographystyle{iclr2027_conference}
\bibliography{iclr2027_conference}


\appendix

\section{Limitations and Scope}
\label{app:limitations}

Our evaluation focuses on representative persistent-memory systems, benchmarks, and the query- and memory-state mutations defined in this work. As with fuzzing more generally, the explored test space is not exhaustive, and the specific failures discovered may vary with memory backend, model, and system configuration. In addition, \system is designed to discover and localize memory-use failures rather than diagnose their internal causes or prescribe mitigation strategies. Extending the framework to additional memory architectures, mutation operators, and deployment settings is a natural direction for future work.

\section{Technical Implementation Details}
\label{app:tech}

\subsection{Mutation Construction and Validation}
\label{app:mutation-implementation}

Given a seed \(x=(M_t,q)\) and a mutation obligation
\(\omega=(x,o,u)\), \system constructs one candidate mutant according to
operator \(o\) and the fixed mutation target \(u\). The information needed
for mutation comes from the mutation descriptor \(P(x)\) and checkpoint
provenance, including the query fields, available replacement targets,
absent targets, and memory operations supported by the backend. The mutation
target and any new value are fixed before execution. When a query requires
free-form rewriting, an LLM may be used to realize the intended change in
natural language, but it does not choose the mutation target or determine
whether the candidate is valid. Each candidate is validated before
execution using the parent query, \(P(x)\), checkpoint provenance, and
observable results of memory operations. The validator never uses the
mutant retrieval \(R_k(x')\), response \(y(x')\), or failure result
\(F_o(x,x')\).

\textbf{Meaning-preserving query mutation.}
For a meaning-preserving mutation, \system keeps \(M_t\) fixed and rewrites
\(q\) while preserving its requested entity, attribute, temporal scope,
answerability, and other answer-affecting constraints. We implement the
four transformations used in the main text: paraphrasing, shortening
removable context, adding nonrestrictive context, and changing syntactic
form. Query fields recorded in \(P(x)\) are treated as protected during
generation. A removable span can be deleted directly, and template-based
syntactic changes are also applied directly when possible; otherwise, the
protected fields and the intended transformation are given to an LLM for
language realization. The resulting \(q'\) is accepted only when the
protected fields remain unchanged and no new condition is introduced that
could alter the answer. For rewrites whose wording changes substantially,
we additionally use a semantic checker as a rejection filter, but its
judgment alone is not sufficient to accept the candidate. Since the query
meaning is preserved, the evaluator keeps \(a'=a\) and the same valid
support for the mutant.

\begin{casebox}{\bf Meaning-Preserving Query Mutation}
Parent: ``Based on the information stored earlier, where does Emma live?''

Mutant: ``Where does Emma live?''

The surrounding phrase is removed, while the requested entity and attribute
remain unchanged.
\end{casebox}

\textbf{Target-changing query mutation.}
A target-changing mutation also keeps \(M_t\) fixed, but changes one
requested entity, attribute, or temporal scope. If \(z\) is the selected
query field, its replacement \(z'\) is chosen from the valid replacement
targets recorded in \(P(x)\), and must correspond to information supported
by the same checkpoint. When the target appears as a replaceable span,
\system applies \(q'=q[z\leftarrow z']\) directly. If this substitution
produces an unnatural or ungrammatical query, an LLM rewrites the sentence
using the fixed replacement \(z'\) while all other query fields remain
protected. Validation checks that the designated field changes to \(z'\)
and that the remaining query fields are unchanged. Once the replacement is
fixed, the evaluator constructs \(G(x')\) from its checkpoint provenance;
this reference is not exposed to the fuzzing search.

\begin{casebox}{\bf Target-Changing Query Mutation}
Suppose \(M_t\) contains supported residence information for both Emma and
Bob.

Parent: ``Where does Emma live?''

Mutant: ``Where does Bob live?''

Only the requested entity changes, and the reference for Bob is constructed
from the same checkpoint.
\end{casebox}

\textbf{Unsupported query mutation.}
For an unsupported mutation, \system replaces one requested target with an
absent target \(z_{\emptyset}\) recorded in \(P(x)\). The replacement must
be type-compatible with the original query so that the mutant remains a
natural request, and we restrict these mutations to memory-exclusive
targets whose answers concern synthetic or session-local information rather
than public facts that may already be known by the base LLM. As with
target-changing mutations, direct slot replacement is used when possible
and an LLM is used only to produce a grammatical realization when needed.
Validation checks the checkpoint provenance before execution and accepts the
candidate only when no memory record supports the requested
entity-attribute-scope combination. Absence is never inferred from a
failed retrieval or an incorrect response. For an accepted mutant, the
evaluator sets \(a'=\bot\).

\begin{casebox}{\bf Unsupported Query Mutation}
Suppose the session contains information about the synthetic project
\emph{Orion}, including its owner and current status, but no budget code.

Parent: ``Who owns Project Orion?''

Mutant: ``What is Project Orion's budget code?''

The mutant remains a valid question about the same session-local entity, but
the requested information is absent from \(M_t\).
\end{casebox}

\textbf{Update memory-state mutation.}
An update keeps \(q\) fixed and changes the value of an existing
entity-attribute pair in an isolated copy of \(M_t\). The mutation target
\(u\) identifies the existing fact and its current value
\(v_{\mathrm{old}}\); before execution, \system chooses a type-compatible
value \(v_{\mathrm{new}}\neq v_{\mathrm{old}}\). If the backend provides an
explicit update operation, \system applies it directly. If updates are
represented through later interactions, \system appends a new interaction
expressing \(v_{\mathrm{new}}\) to the history that produced \(M_t\) and
lets the memory system process the update normally, producing \(M_t'\).
Validation checks that the new value concerns the same entity and attribute
and that the backend records it as the later or active value according to
its own update semantics. Physical removal of \(v_{\mathrm{old}}\) is not
required, since some systems retain historical records. For evaluation,
\(a'\) is changed to \(v_{\mathrm{new}}\), while evidence for
\(v_{\mathrm{old}}\) is placed in \(E'^{-}\).

\begin{casebox}{\bf Update Memory-State Mutation}
Parent memory: ``Emma lives in Chicago.''

Update: ``Emma now lives in Seattle.''

The query remains ``Where does Emma live?'' The updated state \(M_t'\)
should treat Seattle as the current value, while the earlier Chicago record
may remain as history.
\end{casebox}

\textbf{Deletion memory-state mutation.}
A deletion keeps \(q\) fixed and removes a selected fact from an isolated
copy of \(M_t\). The target \(u\) is identified from checkpoint provenance
and, when available, by its backend memory identifier. \system invokes the
backend's native delete or forget operation when one exists; if deletion is
expressed through the system's normal interaction interface, the
corresponding operation is issued in a separate session. When several
records support the same fact, the mutation obligation specifies which
support is removed, and all supporting records are removed when the intended
mutant requires the fact to become unsupported. Validation requires that the
target exists before the operation and is removed or marked inactive
afterward; when the backend returns an explicit operation status, the
operation must also report success. The deleted evidence is placed in
\(E'^{-}\), and the evaluator sets \(a'=\bot\) when no valid support for the
answer remains.

\begin{casebox}{\bf Deletion Memory-State Mutation}
Parent memory: ``Emma lives in Chicago.''

Deletion: remove the stored residence record for Emma.

If no other residence record remains, the fixed query ``Where does Emma
live?'' has \(a'=\bot\) in the mutated state.
\end{casebox}

\textbf{Unrelated memory-state mutation.}
An unrelated change keeps \(q\) and its supporting memory unchanged while
modifying another part of the memory state. Using checkpoint provenance,
\system selects an entity-attribute pair that is outside the target of
\(q\), or creates a new fact that is unrelated to the requested
information, and applies the change through the backend's normal write or
update operation. Validation checks that the modified record does not share
the query target and that the evidence supporting the fixed query remains
unchanged between \(M_t\) and \(M_t'\). Because the information needed by
\(q\) is untouched, the evaluator keeps \(G(x')=G(x)\). This mutation acts
as a control for whether an irrelevant memory change unexpectedly alters
memory use for the same query.

\begin{casebox}{\bf Unrelated Memory-State Mutation}
Fixed query: ``Where does Emma live?''

Unrelated change: add ``Bob prefers skiing.''

Emma's residence information is unchanged, so the expected answer and valid
support remain the same after the mutation.
\end{casebox}

\textbf{General mutation validation.}
Before execution, every generated mutant is passed through the same
validation pipeline. \system first checks that the mutant is well formed and
that the intended change can be identified unambiguously. For query
mutations, we rerun the same field extractor used to build \(P(x)\) on the
generated query \(q'\), and compare the resulting entity, attribute,
temporal scope, and answerability with the operator requirement. A
meaning-preserving mutant is kept only when these fields remain unchanged; a
target-changing mutant is kept only when the designated target changes to
the selected replacement and the remaining fields stay fixed; an
unsupported mutant is kept only when the selected memory-exclusive target
is still absent from checkpoint provenance. Candidates with malformed text,
multiple unintended field changes, ambiguous references, or a target that
cannot be verified are discarded.

For memory-state mutations, \system validates the requested change by
comparing the selected memory item before and after the operation, or by
using the backend's operation result when direct state inspection is
available. An update must record \(v_{\mathrm{new}}\) for the same
entity-attribute pair targeted by the obligation; a deletion must remove
or deactivate the selected support; and an unrelated change must leave the
query-relevant evidence unchanged. We write \(V_o(x,x')=1\) when all checks
required by operator \(o\) pass, and \(V_o(x,x')=0\) otherwise. Only
candidates with \(V_o(x,x')=1\) are executed. These checks use the generated
query, checkpoint provenance, memory contents, and memory-operation logs,
but never the mutant retrieval \(R_k(x')\), response \(y(x')\), or failure
result \(F_o(x,x')\).

\textbf{Validation with hidden memory states.}
Some tested systems do not expose their stored entries directly. For these
systems, \system only treats a memory-state mutation as validated when its
completion can be checked independently of the test response, such as
through a successful memory API call, an operation log, or a
memory-management interface showing the resulting state. We do not validate
a mutation by asking the same agent whether it remembers the new or deleted
fact, since that would use the behavior under test to establish the validity
of the test itself. If the requested state change cannot be verified
independently, the candidate is not counted as a validated memory-state
test.

Across all operators, validation is completed before fuzzing execution. A
candidate that fails its operator-specific checks is discarded and its
mutation obligation remains uncovered. Thus, whether a mutant is considered
valid depends only on the intended query or memory change, not on whether
the resulting execution later exposes a memory-use failure.
\subsection{Implementation Detail of Seed Retention and Iteration}
\label{app:seed-retention}

\system uses observed execution behavior to decide which mutants should
continue to later fuzzing rounds. Failure results \(F_o\) are stored only for
evaluation and never affect retention or prioritization. For systems with
visible retrieval, \system uses three feedback indicators:
\textbf{retrieval coverage}, \textbf{retrieval novelty}, and
\textbf{retrieval divergence}. The first two determine whether a mutant is
retained, while divergence further prioritizes retained meaning-preserving
mutants.

\textbf{Retrieval coverage.}
Let \(R_k(x')=[m_1,\ldots,m_k]\) denote the top-\(k\) memories retrieved by
mutant \(x'\). We map each memory \(m\) to a stable identifier \(\rho(m)\),
using its entity--attribute pair when available and its backend memory
identifier otherwise. The retrieved entries are
\(\mathrm{Reg}(x')=\{\rho(m)\mid m\in R_k(x')\}\). Let
\(\mathcal{V}_r\) contain the memory entries reached earlier in round \(r\).
We define the retrieval-coverage gain as
\[
R_{\mathrm{cov}}(x')
=
\frac{
|\mathrm{Reg}(x')\setminus\mathcal{V}_r|
}{
\max(1,|\mathrm{Reg}(x')|)
}
\]
A positive \(R_{\mathrm{cov}}(x')\) means that the mutant reaches memory
that has not been explored earlier in the round.

\textbf{Retrieval novelty.}
Coverage records which entries have been reached, but not how they are
selected and ranked. We therefore represent the ordered retrieval result by
\(\kappa_R(x')=(\rho(m_1),\ldots,\rho(m_k))\). Let
\(\mathcal{K}_r\) contain the top-\(k\) signatures observed earlier in round
\(r\), and define \(n(x')=1\) when
\(\kappa_R(x')\notin\mathcal{K}_r\), and \(0\) otherwise. Thus, retrieval
novelty captures a new top-\(k\) retrieval pattern, even when its individual
memory entries have already been reached. Following the retention rule in
\S\ref{subsec:execution}, a mutant is retained when
\(R_{\mathrm{cov}}(x')>0\) or \(n(x')=1\).

\textbf{Retrieval divergence.}
For meaning-preserving mutations, \system additionally measures how much the
retrieval ranking changes from the parent. We use Rank-Biased Overlap (RBO),
which places more weight on agreement near the top of two ranked lists. For
lists \(S\) and \(T\), let
\(A_d=|S_{1:d}\cap T_{1:d}|/d\). We use
\(\mathrm{RBO}_{p}(S,T)=(1-p)\sum_{d=1}^{k}p^{d-1}A_d+p^kA_k\)
with \(p=0.9\), and define
\[
R_{\mathrm{div}}(x,x')
=
1-\mathrm{RBO}_{0.9}(R_k(x),R_k(x'))
\]
A larger value indicates a larger change near the top of the retrieval
ranking. Divergence is used only to prioritize meaning-preserving mutants;
it is not an additional retention condition.

Mutation-obligation coverage is handled before this retention priority:
uncovered obligations receive generation attempts before extra variants are
assigned to obligations that have already been exercised. When multiple
retained mutants compete for the remaining execution budget, \system
combines the three retrieval-feedback indicators above. Let
\(\mathbb{1}_{\mathrm{mp}}(x')=1\) when \(x'\) is produced by a
meaning-preserving mutation and \(0\) otherwise. We prioritize retained
mutants by
\begin{equation}
s(x')
=
\frac{
R_{\mathrm{cov}}(x')
+n(x')
+\mathbb{1}_{\mathrm{mp}}(x')R_{\mathrm{div}}(x,x')
}{
2+\mathbb{1}_{\mathrm{mp}}(x')
}
\label{eq:seed-priority}
\end{equation}
which gives equal weight to the available feedback indicators.
After each execution, \(\mathcal{V}_r\) and \(\mathcal{K}_r\) are updated
before the next mutant is considered.

\textbf{Hidden retrieval.}
When retrieval is hidden, \system cannot observe retrieval coverage,
retrieval novelty, or retrieval divergence. It therefore retains mutants
based on previously unseen response behavior, following
\S\ref{subsec:execution}. Each response \(y(x')\) is encoded with a fixed
sentence encoder and compared with response clusters already observed in the
round. If its maximum cosine similarity to the existing clusters is below a
fixed threshold \(\tau\), it forms a new cluster and the mutant is retained.
The encoder and \(\tau\) are fixed before fuzzing.

Let \(\mathcal{R}_r\) denote the mutants retained in round \(r\). The next
seed corpus is
\(\mathcal{S}_{r+1}=\mathcal{S}_r\cup\mathcal{R}_r\). Each retained mutant
is then treated as an ordinary seed in the next round: its current memory
state takes the role of \(M_t\), and either its query or memory state may be
mutated again. Failure results \(F_o\) remain separate throughout this
process.
\section{Mutation Case Studies}
\label{app:mutation-cases}

We provide concrete examples of the six mutation operators used by
\system. The query-mutation examples are constructed from benchmark
questions and references in LoCoMo~\citep{maharana2024evaluating} and
LongMemEval-S~\citep{ICLR2025_d813d324}. For memory-state mutations, we
start from benchmark-derived memory states and apply the valid state
transition defined by the corresponding operator. These examples illustrate
the parent--mutant relation and the expected behavior checked by \system;
they do not assume that every mutant necessarily triggers a failure on every
memory backend.

\begin{casebox2}{Case Study 1: Meaning-Preserving Query Mutation}
\textbf{A surface reformulation should preserve the same memory use.}
LoCoMo contains the query
\emph{``What are Melanie's pets' names?''}, with the expected answer
\emph{Oliver, Luna, Bailey}. \system rewrites it as
\emph{``What are the names of Melanie's pets?''} while preserving the
requested entity, attribute, temporal scope, and answerability. The
parent--mutant pair is
\((M,q)\rightarrow(M,q')\), and the reference remains unchanged,
\(G(M,q')=G(M,q)\). A failure is exposed if the reformulation causes the
agent to lose the supporting pet memories or return an answer inconsistent
with the original benchmark-supported answer.
\end{casebox2}

\begin{casebox2}{Case Study 2: Target-Changing Query Mutation}
\textbf{Changing one supported target should redirect memory use to the new
target.}
LoCoMo asks
\emph{``When did Melanie go to the museum?''}, whose expected answer is
\emph{5 July 2023}. The same conversation also supports
\emph{``When did Melanie go to the pottery workshop?''}, with the expected
answer \emph{the Friday before 15 July 2023}. \system therefore changes the
event target from \emph{museum} to \emph{pottery workshop} while preserving
the entity and requested temporal relation. Unlike the meaning-preserving
case, the mutant has a new reference derived independently from the same
checkpoint. A failure is exposed if the system continues to use the museum
evidence, returns the museum date, or otherwise fails to switch to the
evidence supporting the new target.
\end{casebox2}

\begin{casebox2}{Case Study 3: Unsupported Query Mutation}
\textbf{Replacing a supported target with an unsupported target should
remove answerability.}
LongMemEval-S contains the supported query
\emph{``Which project did I start first, the Ferrari model or the Japanese
Zero fighter plane model?''}, whose expected answer is
\emph{the Japanese Zero fighter plane model}. A corresponding abstention
case replaces the second target with the
\emph{Porsche 991 Turbo S model}, which is not supported as a started
project by the available history. This gives the mutation
\(q[z\leftarrow z_{\emptyset}]\) with \(a'=\bot\).
The lack of support is established from checkpoint provenance before the
mutant is executed. A failure is exposed if the agent nevertheless returns
a definite memory-backed ordering instead of recognizing that the available
memory is insufficient.
\end{casebox2}

\begin{casebox2}{Case Study 4: Update Mutation}
\textbf{A later valid value should replace the earlier value used for the
same query.}
A LongMemEval-S knowledge-update instance records a change in the user's
tennis schedule at the local park: the earlier state says that the user plays
with friends \emph{every week on Sunday}, while the later state says
\emph{every other week on Sunday}. \system keeps the query
\emph{``How often do I play tennis with my friends at the local park?''}
fixed and changes only the memory state,
\((M,q)\rightarrow(M',q)\). The later interaction is supplied to the tested
memory system, which constructs \(M'\) through its native update behavior;
\system does not manually rewrite the resulting memory. After the update,
the expected answer is the new frequency, while evidence supporting the
weekly schedule becomes stale evidence in \(E^{-}\). Returning
\emph{every week} after the update exposes a stale-memory failure.
\end{casebox2}

\begin{casebox2}{Case Study 5: Deletion Mutation}
\textbf{Deleted information should no longer support the fixed query.}
LoCoMo contains the query
\emph{``What did Caroline research?''}, with the expected answer
\emph{Adoption agencies}. Starting from a checkpoint in which the
corresponding fact is stored, \system removes the memory entry supporting
that fact while keeping the query fixed:
\((M,q)\rightarrow(M',q)\). The deletion is performed through the tested
backend's native memory operation and is independently verified before the
mutant is executed. If no other valid support remains, the post-deletion
reference becomes \(a'=\bot\), and the removed evidence is placed in
\(E^{-}\). Retrieving the deleted evidence or continuing to answer
\emph{Adoption agencies} after the deletion constitutes a deletion failure.
\end{casebox2}

\begin{casebox2}{Case Study 6: Unrelated-Change Mutation}
\textbf{An unrelated memory change should not perturb the result of a fixed
query.}
LoCoMo asks
\emph{``What instruments does Melanie play?''}, with the expected answer
\emph{clarinet and violin}. The same conversation contains unrelated
information about Caroline's adoption process. \system introduces a valid
write concerning this Caroline-related information while leaving the query
about Melanie unchanged. Because the new memory concerns a different entity
and attribute, the query reference must remain invariant:
\(G(M',q)=G(M,q)\). The backend is free to incorporate the new information
into its memory state, but the evidence and answer used for Melanie's
instruments should remain unaffected. If the unrelated write changes the
selected support or alters the final answer, \system reports a
non-interference failure.
\end{casebox2}

\end{document}